\documentclass[showkeys,pra,aps,10pt]{revtex4-1}

\begin{document}

%\markboth{Fabris et al}
%{Acoustic black holes in neo-Newtonian theory}

%%%%%%%%%%%%%%%%%%%%% Publisher's Area please ignore %%%%%%%%%%%%%%
%\catchline{}{}{}{}{}
%%%%%%%%%%%%%%%%%%%%%%%%%%%%%%%%%%%%%%%%%%%%%%%%%%%%%%%%%%%%%%%%%%%

\title{A note on acoustic black holes in neo-Newtonian theory
}

\author{\footnotesize J.~C.~Fabris\footnote{Email: fabris@pq.cnpq.br}\,, O.~F.~Piattella\footnote{Email: oliver.piattella@pq.cnpq.br}\,, H.~E.~S.~Velten\footnote{Email: velten@cce.ufes.br}  }

\affiliation{Universidade Federal do Esp\'{\i}rito Santo (UFES)\\
Vit\'oria, ES - Brazil}

\author{I.~G.~Salako\footnote{Email: ines.salako@imsp-uac.org}\,, J.~Tossa\footnote{Email: joel.tossa@imsp-uac.org}}

\affiliation{Institut de Math\'ematiques et de Sciences Physiques (IMSP)\\
Porto-Novo, B\'enin}

%\pub{Received (Day Month Year)}{Revised (Day Month Year)}

\begin{abstract}
Newtonian fluid dynamics allows the construction of acoustic metrics from which black hole configurations can be studied. However, relativistic pressure effects are neglected within Newtonian theory. We study acoustic black holes in the framework of neo-Newtonian hydrodynamics, which is designed to take into account relativistic inertial effects of the pressure $p$. Within this new hydrodynamical context we show how $p$ can influence the formation of the acoustic horizons.

\end{abstract}
\maketitle

%\ccode{PACS Nos.:  04.70.Bw, 47.50.Gj}

\section{Introduction}

One of the most spectacular predictions of General Relativity (GR) are black holes, i.e. collapsed objects that have a singularity hidden by an event horizon. Such objects, without a counterpart in Newtonian theory of gravity, give rise to many interesting effects, from astrophysics to elementary particle physics. From the astrophysics point of view, the possible presence of a black hole in the galactic center may have impact on the dynamics and stability of the whole galaxy. From the fundamental point of view, on the other hand, the existence of quantum fields in the space-time determined by a black hole may cause its evaporation via particle creation \cite{fcreation}.

Even if there are many indirect evidences for the existence of black holes, mainly at astrophysical level, they have until now escaped a direct detection. One of the main reason is that the concept of black hole is connected with the notion of an event horizon, i.e. a surface which hides an interior region from which nothing can escape. This fact implies that essentially all possible evidences for the existence of a black hole come from indirect observations.

More recently, it has been argued that the essential of the physics of a black hole can be obtained from some ordinary systems in a laboratory. One example is a fluid flowing supersonically into a sink, leading to the appearance of a kind of event horizon, now concerning the propagation of sound waves. This possibility has led to the appearance of the so-called analogous systems, i.e. ordinary systems that mimic the properties typical of a black hole. For a review, see Ref.~\cite{M1}.

The main point of the study of black hole physics via analogous systems is that, under certain conditions, the hydrodynamics of an ordinary fluid can be represented by equations similar to those of a scalar field in the black hole space-time, with the emergence of a kind of "effective metric". Hence, the propagation of waves (or phonons, depending on the analogue system used) obeys equations similar to those of particles and light in the vicinity of a black hole. In this sense,
it is possible to speak of horizon and geodesics, and, as far as we can trust in the analogy between the gravitational and hydrodynamical systems, the characteristic effects of the black hole, mainly the presence of a horizon and connected phenomena, can be studied in the laboratory.

In general, analogous systems employ a classical, Newtonian treatment, even if some quantum systems are also considered. Of course, a black hole is a relativistic gravitational phenomenon. Indeed, in the relativistic context the inertial and gravitational effects of the pressure are essential for a correct description of the system. In some cases, relativistic pressure effects can be incorporated in a Newtonian framework, even if approximatively. This is the case of the neo-Newtonian theory, a modification of the usual Newtonian theory incorporating adequately the pressure into the dynamics. Our main goal in this paper is to address the following question: can this neo-Newtonian framework be relevant to the analogous system program? We will see that performing a neo-Newtonian analysis, the pressure of the medium enters directly in the expressions and in some circumstances it is possible to diminish considerably the speed of the flowing fluid in order to generate the acoustic horizon.

In Sec.~\ref{Sec:NewtAc}, we revise the Newtonian formulation for the acoustic black hole. Then, in Sec.~\ref{Sec:NeoNewtAc} we introduce the neo-Newtonian framework and calculate the acoustic metric. In Sec. 4 we conclude by making some considerations on the possibility of obtaining systems with desirable features in the laboratory.

\section{Acoustic black hole in Newtonian theory}\label{Sec:NewtAc}

The basic equations of Newtonian hydrodynamics for an inviscid perfect fluid are the following \cite{M1,e1,e2,e3,e4}:
\begin{eqnarray}
\partial_{t} \rho + \nabla\cdot(\rho\vec{v})  &=& 0, \label{eq100'}\\
\rho \frac{d \vec{v}}{dt} \equiv  \rho [\partial_{t}\vec{v}+
(\vec{v}\cdot\nabla)\vec{v} ] &=& -\nabla p, \label{eq200'}
%\nabla^{2} \Phi &=& 4G \pi \rho.      \label{eq300'}
\end{eqnarray}
where $\rho$ is the fluid density, $p$ its pressure and $\vec v$ its the velocity field. Equation~(\ref{eq200'}) can be rewritten using the equality,
\begin{eqnarray}
  (\vec{v}\cdot\nabla)\vec{v} = (\nabla\times\vec{v})\times \vec{v} + \nabla\left(\frac{v^{2}}{2}\right)\;,   \label{eq4''}
\end{eqnarray}
leading to
\begin{eqnarray}
 \partial_{t}\vec{v}= (-\nabla\times\vec{v})\times \vec{v} - \nabla\left(\frac{v^{2}}{2}\right) - \frac{1}{\rho}\nabla p.
 \label{eq5}
\end{eqnarray}
Let us consider the fluid to be irrotational, i.e. $\vec{v} = -\nabla \phi$, where $\phi$ is the velocity potential. Assuming the fluid to be also barotropic, that means $p = p(\rho)$, it is possible to define the specific enthalpy
\begin{equation}
 h(p) = \int_0^p\frac{dp'}{\rho(p')}\;,
\end{equation}
as a function of the pressure only and then one can write
\begin{equation}
 \nabla h = \frac{1}{\rho} \nabla p\;,
\end{equation}
with which Eq.~(\ref{eq5}) becomes
\begin{eqnarray}
- \partial_{t}\phi + \frac{1}{2} (\nabla \phi)^{2} + h = 0,  \label{eq6''}
\end{eqnarray}
which is Bernoulli's equation. These are the fundamental equations to be used.

\subsection{Fluctuations}

In order to construct the effective geometry from the fluid configuration described above, it is necessary to
consider fluctuations on a given background. Hence, we must linearise equations  (\ref{eq100'}) and (\ref{eq6''}) by perturbing $\rho, \vec v$ and $\phi$:
\begin{eqnarray}
 \rho= \rho_{0} + \varepsilon \rho_{1}+ 0(\varepsilon^{2}), \quad
 \vec v = \vec v_{0} + \varepsilon \vec v_{1} + 0(\varepsilon^{2}), \quad
\phi= \phi_{0} + \varepsilon \phi_{1}+ 0(\varepsilon^{2}). \label{eq73'}
\end{eqnarray}
The linearised continuity equation becomes,
 \begin{eqnarray}
\partial_{t} \rho_{0} + \nabla.(\rho_{0} \vec{v_{0}})= 0,  \\
 \partial_{t} \rho_{1} + \nabla.( \rho_{1} \vec{v_{0}} + \rho_{0} \vec{v_{1}})= 0.    \label{eq8'}
 \end{eqnarray}
Developing the enthalpy relation around $p_{0}$ gives,
\begin{eqnarray}
 h = h_{0} + h'(p_{0})(p-p_{0}) +  0(\varepsilon^{2}) = h_{0} + \frac{1}{\rho_{0}}(p-p_{0})+  0(\varepsilon^{2})= h_{0}  \label{eq9'}
 +\varepsilon \frac{p_{1}}{\rho_{0}}.
\end{eqnarray}
Inserting (\ref{eq73'}) and (\ref{eq9'}) in (\ref{eq6''}), we find
\begin{eqnarray}
p_{1}= \rho_{0} (\partial_{t}\phi_{1} +\vec{v_{0}}.\nabla \phi_{1}).   \label{eq10'}
\end{eqnarray}
Since
\begin{equation}
 \rho_{1} = \frac{\partial {\rho}}{\partial {p}} p_{1}
 = \frac{\partial {\rho}}{\partial {p}} \rho_{0} (\partial_{t}\phi_{1}
 +\vec{v_{0}}.\nabla \phi_{1}),
 \end{equation}
equation (\ref{eq8'}) becomes
\begin{eqnarray}
 -\partial_{t} \Big[\frac{\partial {\rho}}{\partial {p}}\rho_{0}(\partial_{t}\phi_{1} +\vec{v_{0}}.\nabla \phi_{1} )\Big] &+& \nabla\cdot\Big[-\frac{\partial {\rho}}{\partial {p}}\rho_{0} \vec{v_{0}}(\partial_{t}\phi_{1} +\vec{v_{0}}.\nabla \phi_{1})\Big] \nonumber\\
 &+& \nabla\cdot\Big(\rho_{0} \nabla \phi_{1}\Big)=0, \label{eq11'}
\end{eqnarray}
where one recognises the local speed of sound $c_{s}^{2}=\frac{\partial {p}}{\partial {\rho}} $. Equation~(\ref{eq11'}), introducing $(3 + 1)$-dimensional space-time coordinates, takes the form
\begin{eqnarray}
 \partial_{\mu} (f^{\mu\nu} \partial_{\nu} \phi_{1}) = 0\;,   \label{eq13'}
\end{eqnarray}
with
\begin{equation}
{ f^{\mu\nu}(t,\vec{x}}) =\frac{\rho_{0}}{c_{s}^{2}} \left(
\begin{array}{ccc}
 -1 & -v^{j}_{0}  \\
 -v^{i}_{0}   & c_{s}^{2}\delta^{i}_{j} -v^{i}_{0}v^{j}_{0}
\end{array}
\right).
\end{equation}

\subsection{Acoustic metric}

For a barotropic, non viscous and locally irrotational fluid, the equation of  motion for the velocity potential is identical to the D'Alembertian of a massless scalar field. The Klein-Gordon equation \cite{M} for a massless scalar field is,
\begin{eqnarray}
 \frac{1}{\sqrt{-g}} \partial_{\mu} ( \sqrt{-g} g^{\mu\nu} \partial_{\nu} \phi_{1}) = 0.   \label{eq14'}
\end{eqnarray}
Therefore, identifying $f^{\mu\nu}=\sqrt{-g} g^{\mu\nu}$, one obtains
\begin{equation}
 \det f^{\mu\nu}= g\;, \quad g= -\frac{\rho^{4}_{0}}{c_{s}^{2}}\;, \quad \sqrt{-g} =  \frac{\rho^{2}_{0}}{c_{s}}\;,
\end{equation}
and we can read off the contravariant acoustic metric:
\begin{equation}
{  g^{\mu\nu}(t,\vec{x}}) = \frac{1}{\rho_{0} c_{s}} \left(
\begin{array}{ccc}
 -1  & -v^{j}_{0}  \\
 -v^{i}_{0} & c_{s}^{2}\delta^{i}_{j} -v^{i}_{0}v^{j}_{0}
\end{array}
\right).
\end{equation}
The covariant acoustic metric is then the following:
\begin{equation}
{ g_{\mu\nu}(t,\vec{x}}) =   \frac{\rho_{0}}{c_{s}}\left(
\begin{array}{ccc}
 -(c_{s}^{2}-v^{2}_{0}) &  -v^{j}_{0}  \\
 -v^{i}_{0} & \delta_{ij}
\end{array}
\right),
\end{equation}
which, written as line element, is
\begin{eqnarray}
 ds^{2} =  \frac{\rho_{0}}{c_{s}} \Big[-c_{s}^{2}dt^{2}+
 (dx^{i}-v^{i}_{0}dt)\delta_{ij}(dx^{j}-v^{j}_{0}dt)\Big].    \label{eq15'}
\end{eqnarray}

\subsection{Ergo-region and event horizon}

In order to introduce the concepts of ergo-region and event horizon for acoustic black holes, it is necessary to consider a configuration with rotation. The flow model adapted to that is the draining bathtub proposed by Visser \cite{M1}. The fluid velocity must be expressed in terms of polar coordinates. From the continuity equation, assuming that $\rho$ is position-independent, we find the radial component of the velocity:
\begin{eqnarray}
 \partial_{r} (\rho r v_{r}) = 0\;, \qquad \Rightarrow \qquad v_{r} = \frac{A}{r}\;, \label{eq16'}
\end{eqnarray}
where $A$ is a constant. Considering that the fluid is locally irrotational, we find its angular velocity component:
\begin{eqnarray}
 \nabla \times \vec{v} = 0\;, \qquad \Rightarrow \qquad v_{\theta} = \frac{B}{r}\;, \label{eq17'}
\end{eqnarray}
where $B$ is another constant. The velocity can be thus written as
\begin{eqnarray}
 \vec{v}= \frac{A}{r} \hat{r} + \frac{B}{r} \hat{\theta}\;,  \label{eq17'''}
\end{eqnarray}
and the potential becomes
\begin{eqnarray}
 \vec{v} = -\nabla\phi\;, \qquad \Rightarrow \qquad -\phi(r,\theta)= -A \ln r - B \theta\;.    \label{eq18'''}
\end{eqnarray}
Considering (\ref{eq17'''}) and  (\ref{eq18'''}), the acoustic metric
(\ref{eq15'})  becomes
\begin{eqnarray}
 ds^{2} = \frac{\rho_{0}}{c_{s}}  \Big[-c_{s}^{2}dt^{2}+ \left(dr -\frac{A}{r}  dt\right)^{2} + \left(rd\theta -\frac{B}{r}
 dt\right)^{2} + dz^2\Big],   \label{eq19'}
\end{eqnarray}
The properties of this metric become clearer in slightly different
coordinates, defined through the transformations of the time and the azimuthal angle coordinates as,
\begin{eqnarray}
 dt= \bar{dt} +  \frac{A r}{c_{s}^{2}r^{2}-A^{2}}dr\;, \qquad  d\theta= \bar{d\theta} +  \frac{A Bdr}{r(c_{s}^{2}r^{2}-A^{2})}.  \label{eq21'}
\end{eqnarray}
With these new variables the metric becomes, after a rescaling of the time coordinate by $c_s$,
\begin{eqnarray}
 ds^{2} &=&  \frac{\rho_{0}}{c_{s}}\Big[-\left(1 - \frac{A^{2}+B^{2} }{c_{s}^{2}r^{2}}\right)\bar{dt}^{2} + \left(1- \frac{A^{2}}{c_{s}^{2}r^{2}}\right)^{-1}dr^{2}\nonumber\\
 &-&  \frac{2B}{c_{s}} \bar{d\theta} \bar{dt} + r^{2} \bar{d\theta}^{2} + dz^2\Big].    \label{eq22'''}
\end{eqnarray}
From (\ref{eq22'''}), one can calculate the radius of the event horizon and of the ergo-region of the acoustic black hole in the ordinary Newtonian theory:
  \begin{eqnarray}
 1/g_{rr} = 0     \quad \Longrightarrow    r_{\rm h} = \frac{|A|}{c_{s}}    \label{eq23'}
\end{eqnarray}
and
\begin{eqnarray}
  g_{00} = 0    \quad \Longrightarrow     r_{\rm e}= \frac{\sqrt{A^{2}+B^{2}}}{c_{s}}.    \label{eq24'}
\end{eqnarray}
It is important to stress that the event horizon coincides with the notion of apparent horizon for stationary black holes, see Ref. \cite{frolov}.
For $A > 0$ we are dealing with a past event horizon, i.e., acoustic white hole, and for $A < 0$ we are dealing
with a future acoustic horizon, i.e., acoustic black hole.

\section{Neo-Newtonian theory}\label{Sec:NeoNewtAc}

In Refs. \cite{A,B} it has been shown that cosmology can be treated within the context of Newtonian physics. Using the continuity equation and the Euler equation in their traditional form, as in (\ref{eq100'}, \ref{eq200'}), ones shows that, for a universe filled with pressureless matter like in the Einstein-de Sitter model, the main results for the homogeneous, isotropic and expanding geometry in the relativistic context can be reproduced with the Newtonian equations. Of course, this is possible if the gravitational interaction  is included through the Poisson equation. The program established in Refs. \cite{A,B} gave rise to the Newtonian cosmology.

Later, a generalization of this result, when the pressure is not zero, has been determined in \cite{C}. This approach has been refined in Ref. \cite{m2}, leading to the so-called neo-Newtonian theory. In the latter, the fundamental equations are \cite{m2}:
\begin{eqnarray}
 \partial_{t} \rho + \nabla\cdot\left[(\rho+\frac{p}{c^{2}})\vec v\right] &=& 0\;, \label{eq1}\\
 \frac{d \vec{v}}{dt} \equiv  \Big[\partial_{t}\vec{v}+ (\vec{v}\cdot\nabla)\vec{v} \Big] &=&
 -\left(\rho + \frac{p}{c^{2}}\right)^{-1}\nabla p\;, \label{eq2}
% \nabla^{2} \Phi &=& 4 \pi G (\rho+\frac{3p}{c^{2}}).         \label{eq3}
 \end{eqnarray}
where $c$ is the speed of light. When $p/c^2 <<\rho$ we find Eqs.~(\ref{eq100'}, \ref{eq200'}), as expected from the Newtonian formalism. In one sense, the remarkable aspect of the above set of equations is that the inertial mass present in the Newtonian equations is replaced by $\rho + p/c^2$.

We remark that concerning cosmological structure formation, equation (\ref{eq1}) does not reproduce the correct growth of matter density perturbations in an homogeneous, isotropic and expanding background \cite{ademir,rrrr,velten}. Actually, the correct growth is obtained if equation (\ref{eq1}) is modified as
\begin{eqnarray}
\label{cosmo}
 \partial_{t} \rho + \nabla\cdot(\rho\vec v) + \frac{p}{c^{2}}\nabla\cdot\vec{v} &=& 0\;. \label{eq1bis}
 \end{eqnarray}
Such form for the continuity equation leads to the same effective metric as the usual Newtonian case when applied to the fluid configuration considered here, as shown in the appendix. This shows, in a context different from the cosmological one, how the neo-Newtonian formulation is sensible to the the specific symmetries and hypothesis of the problem. This fact may indicate that the construction of a Newtonian counterpart of a relativistic problem
may vary from case to case, deserving a more general, deeper analysis.

Our interest consists in using equations (\ref{eq1}, \ref{eq2}) to describe an acoustic black hole, following the strategy proposed in Refs. \cite{fcreation,fref} in order to test the hypothesis that a black hole may evaporate \cite{z,z1}.

\subsection{Acoustic black holes in neo-Newtonian theory}

Let us assume the equation of state $p = k \rho^{n}$, with $k$ and $n$ constants. Therefore, the fluid is barotropic. We also assume, as in the previous section, the fluid to be inviscid and irrotational. Therefore Eq.~(\ref{eq1}) and the neo-Newtonian version of Eq.~(\ref{eq6''}) become
\begin{eqnarray}
 \partial_{t} \rho + \frac{k}{c^{2}}\nabla\cdot(\rho^{n} \vec{v}) + \nabla\cdot(\rho\vec{v}) &=& 0, \label{eq1''}\\
\partial_{t}\vec v + \vec v\cdot\nabla \vec v +
\frac{\nabla p}{(\rho + \frac{k}{c^{2}} \rho^{n})} &=& 0 \label{eq2''}.
\end{eqnarray}
If $k \ll c^{2}$, both equations reduce to their Newtonian counterpart.

\subsection{Fluctuations}

Let us suppose, as before, a fluid with constant density. Again, we write the fluid velocity as $\vec v = - \nabla\phi$.
We linearise, following the same lines as before, equations (\ref{eq1''}, \ref{eq2''}):
\begin{eqnarray}
 \rho &=& \rho_{0} + \varepsilon \rho_{1} + 0(\varepsilon^{2})\;, \\
 \rho^{n} &=& \left[\rho_{0} + \varepsilon \rho_{1}+ 0(\varepsilon^{2})\right]^{n} \approx
 \rho^{n}_{0} + n \varepsilon \rho^{n-1}_{0} \rho_{1} + ... \;,\\
\phi &=& \phi_{0} + \varepsilon \phi_{1}+ 0(\varepsilon^{2})\;. \label{eq80'}
 \end{eqnarray}
The linearised continuity equation (\ref{eq1''}) becomes,
\begin{eqnarray}
 \partial_{t} \rho_{1} + \gamma \nabla\cdot\left(\rho_{1} \vec{v_{0}}\right) + \gamma' \nabla\cdot\left(\rho_{0} \vec{v_{1}}\right) = 0\;,  \label{eq8'''}
 \end{eqnarray}
where
\begin{equation}
\label{gamma}
\gamma = \left(1+\frac{k n \rho^{n-1}_{0}}{c^{2}}\right) = \left(1 + \frac{c_s^2}{c^2}\right)\;, \qquad  \gamma' = \left(1+\frac{k\rho_0^{n-1} }{c^{2}}\right)\;.
 \end{equation}

 On the other hand, the linearised Euler's equation is,
 \begin{equation}
 \frac{\partial\vec v_1}{\partial t} + \vec v_0\cdot\nabla \vec v_1 + \vec v_1\cdot\nabla \vec v_0 = - \frac{c_s^2}{\gamma'}\frac{\nabla\rho_1}{\rho_0}.
 \end{equation}
 It can be rewritten as
 \begin{equation}
 \frac{\partial \phi_1}{\partial t} + \vec v_0\cdot\nabla\phi_1 = \frac{c_s^2}{\gamma'}\frac{\rho_1}{\rho_0}.
 \end{equation}

Following the same steps as in the previous section, the equation governing the evolution of $\phi_1$ is,
\begin{eqnarray}
 -\partial_{t} \Big[c_{s}^{-2}\rho_{0}(\partial_{t}\phi_{1}
 + \vec{v_{0}}.\nabla \phi_{1} )\Big] \nonumber\\
 +  \nabla\cdot\Big[-\gamma c_{s}^{-2}\rho_{0} \vec{v_{0}}(\partial_{t}\phi_{1} +
  \vec{v_{0}}.\nabla \phi_{1})+ \rho_{0}\nabla \phi_{1}\Big]=0. \label{eq12''}
\end{eqnarray}

Using the commutation of partial derivatives, this equation can be written as,

\begin{eqnarray}
 -\partial_{t} \Big\{c_{s}^{-2}\rho_{0}\Big[\partial_{t}\phi_{1}
 + \Big(\frac{1}{2} + \frac{\gamma}{2}\Big)\vec v_{0}.\nabla \phi_{1}\Big]\Big\} \nonumber\\
 +  \nabla\cdot\Big\{- c_{s}^{-2}\rho_{0} \vec{v_{0}}\Big[\Big(\frac{1}{2} + \frac{\gamma}{2}\Big)\partial_{t}\phi_{1} +
  \gamma\vec{v_{0}}.\nabla \phi_{1}\Big]+ \rho_{0}\nabla \phi_{1}\Big\}=0. \label{eq12bis}
\end{eqnarray}
The effective (acoustic) metric related to this equation is then,
\begin{eqnarray}
 ds'^{2} = \frac{\rho_{0} }{c_{s}}\left[-( c_{s}^{2}-\gamma\;v^{2}_{0} ) dt^{2} -2 v_0 \sigma\, dx\;dt -2 v_0 \sigma\, dy\;dt+ dx^2+dy^2+dz^2 \right]. \label{eq15''}
\end{eqnarray}
 where, from now on, we work with the definition
\begin{equation}
\sigma = \left(\frac{1}{2} +\frac{\gamma}{2}\right).
\end{equation}

\subsection{Ergo-region and event horizon}

Considering again a position-independent density, as in the previous section, one can find again, from the continuity equation, the same velocity field and potential, i.e.,
\begin{eqnarray}
 \vec{v} = \frac{A}{r} \hat r + \frac{B}{r} \hat\theta\;.  \label{eq17''}
\end{eqnarray}
and
\begin{eqnarray}
 \phi(r,\theta) = - A \ln r - B \theta.    \label{eq18'}
\end{eqnarray}
Therefore, considering (\ref{eq17''}) and (\ref{eq18'}), the acoustic metric (\ref{eq15''}) becomes:
\begin{eqnarray}
 ds'^{2} =  -\left[1 - \frac{\gamma \;(A^{2} + B^{2})}{c_s^{2}r^{2}}\right] dt^{2} - \left[1- \frac{(\gamma A)^{2}}{ c_s^{2}r^{2}}\right]^{-1} dr^{2} \nonumber\\ - \frac{2\gamma B}{c_s} d\theta dt + r^{2}d\theta^{2} + dz^2,    \label{eq22'}
\end{eqnarray}
from which we can calculate the radius of the ergo-region and of the event horizon:
\begin{equation}
 r_{\rm e} = \frac{ \sqrt{ \gamma}}{c_s}\;  \sqrt{A^{2}+B^{2}}\;, \qquad r_{\rm h} = \frac{\sqrt{\gamma}}{c_s}\;|A|.
\end{equation}

\subsection{Schwarzschild form of the acoustic metric}

The acoustic metric (\ref{eq15''}) can in principle be written in the form of a Schwarzschild metric. In order to do so, we first write the latter using Painlev\'e-Gullstrand coordinates:
\begin{eqnarray}
 ds^{2}_{\rm S} = -\left(1 - \frac{2GM}{r}\right) dt^{2} \pm \sqrt{\frac{2GM}{r}} drdt + dr^{2} + r^{2}(d\theta^{2} +
\sin^{2}\theta d\phi^{2})\;.  \label{eq44'}
\end{eqnarray}
When compared with the Schwarzschild coordinates, the Painlev\'e-Gullstrand form has the advantage of being regular at the horizon.

Now, we put metric (\ref{eq15''}) in a similar form, by considering the following, purely radial, velocity:
\begin{eqnarray}
 \vec{v}_0 = \sqrt{2GM/r}\hat r\;. \label{eq62}
\end{eqnarray}
With this choice we have a more natural interpretation of the metric in terms of the velocity fo the fluid $\vec{v_0}$.

We use also
\begin{eqnarray}
 \rho(r,t) = \rho(r)\;, \label{eq61}
\end{eqnarray}
which lead to the following form for the continuity equation:
\begin{eqnarray}
 \nabla (\rho \vec{v})= 0\;. \label{eq60''}
\end{eqnarray}
Solving (\ref{eq60''}) we find
\begin{eqnarray}
 \rho \propto r^{-\frac{3}{2}}\;. \label{eq60}
\end{eqnarray}
Taking into account (\ref{eq62}), the acoustic metric (\ref{eq15''}) becomes:
 \begin{eqnarray}
 ds^{2}_{\rm acoustic}  \propto r^{-\frac{3}{2}} \left[-\left(1 - \gamma \;\frac{2GM}{r}\right)dt^{2} \pm \sigma \sqrt{\frac{2GM}{r}} drdt +\right. \nonumber\\ \left. + dr^{2} + r^{2}(d\theta^{2} +
\sin^{2}\theta d\phi^{2})\right]\;.\label{eq60'}
\end{eqnarray}
This is the analogue of the Schwarzschild black hole in the neo-Newtonian theory.

We remark that the Schwarzschild analogous in the ordinary Newtonian theory is recovered if we consider $p$ very small, i.e. in the limit $k \rightarrow 0$ or ($\gamma \rightarrow 1 \;\;and \;\sigma \rightarrow 1 )$.

Metric (\ref{eq60'}) is similar, but not equivalent, to the Painlev\'e - Gullstrand metric presenting an acoustic horizon, i.e. a region where the velocity of the fluid is larger than the speed of sound. In this way, we can consider the Hawking radiation phenomena in this context.

\subsection{Surface gravity and Hawking radiation}

We will now evaluate the surface gravity and the temperature of an acoustic black hole in the context of the neo-Newtonian theory. First, we diagonalize the acoustic metric (\ref{eq15''}) by defining
a new time coordinate:
\begin{eqnarray}
 d\tau = dt +  \frac{\sigma\; \vec{v}_0\cdot\vec{dx}}{c_s^{2}-\gamma \;v^{2}_{0}}\;.
\end{eqnarray}
Therefore, the acoustic metric (\ref{eq15''}) becomes:
\begin{eqnarray}
 ds^{2} = \frac{\rho_{0}}{c_{s}} \Big[-(c_s^{2}-\gamma \;v^{2}_{0})d\tau^{2} +
 \left(\delta_{ij} + \frac{(\sigma^{2}+ \gamma)\; v_{i}v_{j}}{c_s^{2}- \gamma \; v^{2}_{0}}\right)dx^{i}
 dx^{j} \Big].\label{eq71'''}
\end{eqnarray}
In this coordinate system, the absence of time-space cross-terms makes manifest that the acoustic geometry is in fact static (there exists thus a family of space-like hypersurfaces orthogonal to the time-like Killing vector). The surface gravity can be computed from (\ref{eq71'''}) as follows:
\begin{eqnarray}
 \kappa^{2}&=& - \frac{1}{2} \Big[ (\nabla^{\mu}\chi^{\nu}) (\nabla_{\mu}\chi_{\nu})\Big]_{\arrowvert_{\rm horizon} }\cr \label{eq71}
 &=& - \frac{1}{2} \Big[ (g^{\mu \alpha}\nabla_{\alpha}\chi^{\nu}  )(\nabla_{\mu} g_{\nu \rho }\chi^{\rho}) \Big]_{\arrowvert_{\rm horizon} } \cr
 &=&  - \frac{1}{8} \Big[g^{tt}g_{rr}\left(-g^{rr}\frac{d}{dr} g_{tt}\right)^{2}
 + g^{rr}g_{tt}\left(-g^{tt}\frac{d}{dr} g_{tt}\right)^{2} \Big]_{\arrowvert_{\rm horizon}}.
\end{eqnarray}
Hence, we have
\begin{eqnarray}
 \kappa &= & \frac{1}{2 \rho_0} \frac{d}{dr}\left[-g_{tt} \right]_{\arrowvert_{\sqrt{\gamma} \; v_{0}=c_s}} \cr
 &=& \frac{1}{2 \rho_0} \frac{d}{dr}\left[\frac{\rho_0}{c_s} (c_s^{2}-\gamma \;v^{2}_{0})\right]_{\arrowvert_{\sqrt{\gamma} \; v_{0}=c_s}} \cr
&=&  \frac{1}{2 \;c_s} \frac{d}{dr}\left[(c_s^{2}-\gamma \;v^{2}_{0})\right]_{\arrowvert_{\sqrt{\gamma} \; v_{0}=c_s}},  \label{eq201}
\end{eqnarray}
and for the Hawking temperature:
\begin{eqnarray}
 T = \frac{\hbar }{4 \pi c_s k_{B}} \frac{d}{dr}\left[c_s^{2}-\gamma \;v^{2}_{0}\right]_{\arrowvert_{\sqrt{ \gamma } v_{0}=c_s}}\;,  \label{eq202}
\end{eqnarray}
For $k \rightarrow 0$, or ($\gamma \rightarrow 1 \;and\;\sigma \rightarrow 1 $), i.e. the Newtonian limit, relations (\ref{eq201}) and (\ref{eq202})
become
\begin{eqnarray}
 \kappa = \frac{1}{2c_{s}} \frac{d}{dn}\left[c_{s}^{2}-v^{2}_{0}\right]_{\arrowvert_{ v_{0}=c_{s}}}\;,
\end{eqnarray}
 and
\begin{eqnarray}
 T = \frac{\hbar}{4 \pi c_{s} k_{B}} \frac{d}{dn}\left[c_{s}^{2}- v^{2}_{0}\right]_{\arrowvert_{ v_{0}=c_{s}}}.
\end{eqnarray}
Hence, we recover the results for the ordinary Newtonian theory \cite{fref,M1}.

\section{Discussions}

We have carried out in this work part of the analogue gravity program. The possibility of mimicking general relativistic effects, like the horizon notion, via standard Newtonian fluid dynamics opens the window for an experimental observation of such relativistic features.

From cosmology, one learns that a phenomenological description of the cosmic components can be performed with use of the Newtonian hydrodynamics. However, the standard Newtonian approach is valid only for pressureless fluids. Even for cosmological applications, but also for any system in the laboratory, the kinetic pressure of the system is a fundamental quantity which can play a very important role on the dynamics. Moreover, general relativity tell us that pressure is a fundamental component of the energy-momentum tensor. Thus, in order to keep the simplicity of the Newtonian framework and including, at the same time, the relativistic effects of the pressure, we have advocated here the use of the so called neo-Newtonian formalism for the analogue gravity program.

One important aspect of the results reported here is that, perhaps, the specific form of the neo-Newtonian equations, retaining the essential features of
the relativistic problem, may depend on the symmetries of the problem, as a comparison with the cosmological case suggests.

We have constructed acoustic black hole configurations in the context of the neo-Newtonian hydrodynamics. The presence of the parameter $\gamma$ in the effective metric of acoustic black holes in the framework of the neo-Newtonian theory is the main feature of our results. This parameter is always larger than $1$ if $k$ and $n$ have the same sign, that is, if the fluid have the expected positive square speed of sound. Hence, essentially, the velocity on the horizon is diminished by the presence of the pressure. Can this modified acoustic black hole be reproduced in laboratory? If General Relativity is the correct theory of gravitation, the neo-Newtonian theory can be considered as a possible implementation of gravitational effects without leaving completely the Newtonian framework. But, in order to give expressive effects that can be tested in the laboratory is necessary to have a pressure that is negative and, at the same time, can be comparable with density in absolute value.

Bose-Einstein condensates (BECs) provide a useful tool for approaching quantum gravity phenomenology since they are considered candidates for a future experimental observation
of (phononic) Hawking radiation \cite{becHaw} or of particle creation in expanding spacetimes\cite{becpc1,becpc2}. Inded, BECs are very diluted systems. But, since the BEC pressure that can be obtained from the Gross-Pitaevskii equation is proportional to $\rho^{2}$, the necessary configuration may not occur.

Another possibility would be a fluid obeying the Chaplygin gas model, given by $p = - A/\rho$, where $A$ is a constant. In this case, $k$ and $n$ in (\ref{gamma}) have both negative
values implying still $\gamma > 1$. Remark that, despite exhibiting negative pressure, the Chaplygin gas model has positive square speed of sound. Even if nowadays the Chaplygin gas model is evoked in cosmology \cite{j1}, to give a unified scenario for dark matter and dark energy, the Chaplying gas equation of state was considered initially in fluid dynamics and aerodynamics contexts \cite{j2,j3,j4,j5}. If the density can be made small enough, the pressure can be
made very high, and the results described above can be perhaps implemented in laboratory.

\appendix

\section{Alternative form for the neo-Newtonian equations}\label{app}

Let us use now the form for the continuity equation given by (\ref{cosmo}). As already stated, this form of the continuity equation gives the correct
cosmological equations for the background and at linear perturbative level for a constant equation of state.
For the problem treated here, using this form for the continuity equation, we find,

The linearised continuity equation (\ref{cosmo}) becomes,
\begin{eqnarray}
 \partial_{t} \rho_{1} + \nabla\cdot\left(\rho_{1} \vec{v_{0}}\right) + \gamma' \nabla\cdot\left(\rho_{0} \vec{v_{1}}\right) = 0\;.
 \end{eqnarray}
 The linearized Bernoulli equation is the same as before. Hence, we find the following decoupled equation:
\begin{eqnarray}
 -\partial_{t} \Big[c_{s}^{-2}\rho_{0}(\partial_{t}\phi_{1}
 +  \vec{v_{0}}.\nabla \phi_{1} )\Big] \nonumber\\
  + \nabla\cdot\Big[- c_{s}^{-2}\rho_{0} \vec{v_{0}}(\partial_{t}\phi_{1} +
  \vec{v_{0}}.\nabla \phi_{1})+ \rho_{0} \nabla \phi_{1}\Big]=0.
\end{eqnarray}
This is the same equation found in the usual newtonian case.

\section*{Acknowledgments}
 I.G. Salako thank ICTP/IMSP (Benin) for partial financial support. J.C. Fabris, O.F. Piattella and H.E.S. Velten thank CNPq (Brazil) and FAPES (Brazil) for partial financial support. J. Tossa thanks CNPq (Brazil) for partial financial support and the Gravitation and Cosmology Group of UFES for hospitality during
the elaboration of this work. We thank Kirill Bronnikov for this remarks on the text.

\end{document}